\documentclass[a4paper,fleqn,usenatbib]{mnras}

\usepackage[T1]{fontenc}
\usepackage{ae,aecompl}
\pdfoutput=1

\usepackage{graphicx}	
\usepackage{amsmath}	
\usepackage{amssymb}	
\usepackage{hyperref}

\title[Galactic polarized synchrotron simulations for EoR]{Simulations of Galactic polarized synchrotron emission for Epoch of Reionization observations}
\author[M. Spinelli et al.]{M.~Spinelli$^{1}$\thanks{E-mail: mspinelli@uwc.ac.za}, G.~Bernardi$^{2,3,4}$ and M.G.~Santos$^{1,2}$
\\
$^{1}$Department of Physics and Astronomy, University of Western Cape, Cape Town 7535, South Africa\\
$^{2}$SKA SA, 3rd Floor, The Park, Park Road, Pinelands, 7405, South Africa\\
$^{3}$Department of Physics and Electronics, Rhodes University, PO Box 94, Grahamstown, 6140, South Africa\\
$^4$INAF-Istituto di Radioastronomia, via Gobetti 101, 40129, Bologna, Italy
}

\date{Accepted XXX. Received YYY; in original form ZZZ}

\pubyear{2018}

\begin{document}
\label{firstpage}
\pagerange{\pageref{firstpage}--\pageref{lastpage}}
\maketitle

\begin{abstract}
The detection of the redshifted cosmological $21$~cm line signal requires the removal of the Galactic and extragalactic foreground emission, which is orders of magnitude brighter anywhere in the sky. 
Foreground cleaning methods currently used are efficient in removing spectrally smooth components.  However, they struggle in the presence of not spectrally smooth contamination that is, therefore, potentially the most dangerous one. An example of this is the polarized synchrotron emission, which is Faraday rotated by the interstellar medium and leaks into total intensity due to instrumental imperfections. In this work we present new full-sky simulations of this polarized synchrotron emission in the $50-200$~MHz range, obtained from the observed properties of diffuse polarized emission at low frequencies. 
The simulated polarized maps are made publicly available, aiming to provide more realistic templates to simulate the effect of instrumental leakage and the effectiveness of foreground separation techniques.
\end{abstract}



\section{Introduction}\label{intro}

The study of the first luminous sources and the consequent epoch of reionization (EoR) occupies a central place in modern cosmology. Amongst the various probes of this phase of the Universe, the redshifted $21$-cm line is expected to be the most promising one, potentially allowing us to observe even before the first stars started to shine \citep[see][for recent reviews]{Furlanetto2016,McQuinn2016}. 

Measurements of the redshifted 21-cm line are plagued by foregrounds that are a few orders of magnitude brighter than the 21-cm signal anywhere in the sky \citep[e.g.,][]{Bernardi2009,Parsons2014} and can be only separated by leveraging upon their different spectral coherence \citep[e.g.,][]{Santos2005,Dillon2014,Ali2015,Chapman2016,Patil2017, Liu2014a, Liu2014b, Wang2013}. Over the last decade, 21-cm upper limits have steadily improved both from sky-averaged \cite[][]{Bernardi2016,Singh2017,Monsalve2017} and power spectrum \cite[][]{Dillon2015,Jacobs2015,Ali2015,Ewall-Wice2016, Beardsley2016,Patil2017} observations. 

As upper limits become more and more stringent, systematic effects need to be known and modelled more precisely. One of these effects has been recognized since early on to be the leakage from polarized foregrounds into the total intensity where the cosmological signal is measured \citep[][]{Bernardi2010,Jelic2010}. Although usually small (a few percent or less) this leakage can be quite important as 1) the sky polarization can be much larger than the cosmological signal and 2) the polarized signal can have a not smooth frequency dependence which makes this leakage much harder to clean. This frequency dependence is due to Faraday rotation, a rotation of the polarization angle in linearly polarized radiation as it traverses the Galaxy due to its interaction with the Galactic magnetic field (see \citealt{Rybicki&Lightman1986} for further details). 

The most prominent mechanism that leads to polarization leakage into the EoR signal is likely due to the intrinsic polarization response of low frequency receptors (``polarized beams"). \citet{Asad2015,Asad2016} studied the case of EoR observations with the LOFAR telescope that has a relatively narrow field of view ($\sim 6^\circ$ at 150~MHz) and can be pointed to sky regions with relatively faint Galactic polarized emission \citep{Bernardi2010}. Under those specific conditions, they found that the polarization leakage may be kept below the expected 21-cm signal. \citet{Moore2013}, conversely, simulated EoR observations with very wide field of view instruments and found that polarization leakage may be well above the EoR signal, in particular due to the population of polarized extragalactic sources. Current observations have only placed upper limits on the level of all-sky polarized foreground emission \citep{Kohn2016,Moore2017}. 

Recently, \citet{Nunhokee2017} modelled the contamination due to polarization leakage from wide--field polarized beams and found that it is likely to be non-negligible, although its exact magnitude strongly depends upon the properties of all-sky polarized foregrounds. An accurate understanding and modelling of polarized foregrounds is therefore crucial to quantify both the amount of leakage to the 21-cm power spectrum and the effectiveness of foreground separation techniques in the presence of such a leakage.

In this paper we present new all-sky simulations of Galactic polarized emission at frequencies below 200~MHz, aimed at improving the accuracy of polarization leakage simulations. Unlike previous efforts, our simulation do not rely on intensity data but are built from the statistics of observed polarized foregrounds at low frequencies. Moreover, they can be further improved with the inclusion of upcoming observations.

The paper is organized as follows: we give a brief summary of the polarization framework  in section~\ref{theory}, we describe our simulation method in section~\ref{sec:sim};
the simulation results are presented in section~\ref{sec:results} and we conclude in section~\ref{sec:conclusions}.

\section{Theoretical background}\label{theory}

The intensity of the linearly polarized synchrotron emission can be written in a complex form as:
\begin{equation}\label{eq:P}
P=Q+iU=I_Pe^{2i\chi},
\end{equation}
where $I_P=\sqrt{Q^2+U^2}$ is the polarization intensity ($Q$ and $U$ are the standard Stokes parameters) and the polarization angle is:
\begin{equation}\label{eq:phase}
\chi=\frac{1}{2}\arctan\left(U/Q\right).
\end{equation}

As polarized synchrotron emission travels through the interstellar medium (ISM), its polarization angle $\chi$ rotates as a function of the square of the wavelength $\lambda$:
\begin{equation}\label{eq:chi}
\chi \left ( \lambda^2 \right )=\chi_0+\psi \lambda^2
\end{equation}
where $\chi_0$ is the intrinsic polarization angle at the source and $\psi$ is the Faraday depth along the line of sight towards the source \citep[i.e.][]{Burn1966}:
\begin{equation}\label{eq:psi}
 \psi \propto \int_{0}^{\mathrm{source}}d\mathrm{r}~ \mathrm{n_e B_\parallel},
\end{equation}
where $n_e$ is the thermal electron density and $B_\parallel$ is the magnetic field component along the line of sight. The integral is carried out between the observer's location and the source distance. 
 Faraday rotation therefore imprints a specific frequency  $\psi$--dependent coherence on the Stokes $Q$ and $U$ parameters for any given line of sight $\hat{n}$:
\begin{equation*}\label{eq:stokesQ-diffuse}
Q ({\hat{n}, \nu}) = Q_0(\hat{n}) \cos{\left ( 2 \, \psi  \lambda^2 \right )} + U_0(\hat{n}) \sin{\left ( 2 \, \psi  \lambda^2 \right )},
\end{equation*}\vspace{-0.3cm}
\begin{equation}\label{eq:stokesU-diffuse}
U ({\hat{n}, \nu})  =  -U_0(\hat{n}) \sin{\left ( 2 \, \psi  \lambda^2 \right )} + Q_0(\hat{n}) \cos{\left ( 2 \, \psi  \lambda^2 \right )},
\end{equation}
where it is implicit that the frequency $\nu = c/\lambda$ and $Q_0$ and $U_0$ are measured at a given reference frequency, $\nu_0$. Note that the expressions above are only valid for the emission from one single source. The observed synchrotron radiation will be an integral over many emission sources along each line of sight.

The diffuse polarized emission is often analysed using the rotation measure (RM) synthesis technique, presented in \citet{Burn1966} and extended in \citet{Brentjens2005}. This technique is useful  in cases where there is a superposition of emitting regions along the line of sight, with different values of Faraday depth or of faint highly rotating emission. 
In this section,  we briefly outline this technique since this will be helpful to describe 
our simulation strategy in section~\ref{simu_strat}.  For a more detailed review we refer to the original papers or \citet{Heald2009b}. 
The RM synthesis takes advantage of the formal Fourier relation between the polarized emission and the intrinsic polarized flux as a function of Faraday depth.
Indeed, the complex polarized intensity $P$ as a function of wavelength $\lambda^2$ and its Faraday dispersion $\tilde{P}(\psi)$ form a Fourier pair:
\begin{eqnarray}
P(\lambda^2) & = & \int_{-\infty}^{+\infty}\tilde{P}(\psi)e^{2i \psi \lambda^2}d\psi \nonumber \\
\tilde{P}(\psi) & = & \int_{-\infty}^{+\infty}P(\lambda^2)e^{-2i \psi \lambda^2}d\lambda^2,
\label{RM1}
\end{eqnarray}
where $\tilde{P}(\psi)$ can be formally re-written as:
\begin{eqnarray}
\tilde{P}(\psi)=\tilde{Q}(\psi)+i\tilde{U}(\psi), \, \, \, {\rm with} \nonumber \\
I_{\tilde{P}}(\psi)=\sqrt{\tilde{Q}(\psi)^2+\tilde{U}(\psi)^2}. \nonumber
\end{eqnarray}

Since $\lambda^2$ is positive by definition and, in practice, the sampling in $\lambda^2$ space is always incomplete, this formula has been corrected in \citet{Brentjens2005}  and  expressed as:
\begin{equation}\label{eq:RM2}
\tilde{P}_{\mathrm{obs}}(\psi)=K\int_{-\infty}^{+\infty}P_{\mathrm{obs}}(\lambda^2)e^{-2i \psi (\lambda^2-\lambda_0^2)}d\lambda^2=\tilde{P}(\psi) \ast R(\psi)
\end{equation}
where the sampling (or window) function, $W(\lambda^2)$, which sets the frequency range, is included in $P_{\mathrm{obs}}$ and $K$ is the inverse of the integral over this sampling function. The RM transfer function (RMTF):
\begin{equation}\label{eq:RMTF}
R(\psi)=K\int_{-\infty}^{+\infty}W(\lambda^2)e^{-2i \psi (\lambda^2-\lambda_0^2)}d\lambda^2
\end{equation}
determines the resolution in Faraday depth. Note the factor $\lambda_0^2$ (the weighted average of the observed $\lambda^2$), in equation~\ref{eq:RM2} and \ref{eq:RMTF}, that has been introduced to improve the behaviour of the RMTF. 

As with standard Fourier transforms, the FWHM $\delta\psi$ of the main peak of  the RMTF is inversely proportional to the full width of the $\lambda^2$ space covered by observations. The largest scale in $\psi$ space to which one is sensitive is inversely proportional to the shortest wavelength square $\lambda_{\mathrm{min}}^2$, while the maximum observable Faraday depth $\psi_{\mathrm{max}}$ depends on the channel width.

The output of the RM synthesis is a cube of polarized maps, $\tilde{P}(\psi)$ at selected values of Faraday depth $\psi$. 
In the next sections we will model it in order to produce simulated Stokes $Q$ and $U$ maps at the frequencies of interest through equation~\ref{RM1}.

\section{Simulations}\label{sec:sim}
In this section we describe in detail our simulation method for low frequency polarised synchrotron emission. We start by reviewing some of the approaches that can be found in literature in section~\ref{sec:oldlit}, which mostly rely on intensity data at higher frequencies. We then discuss the limitation of these methods and present our simulation recipe based on polarized data in section~\ref{simu_strat}. Section~\ref{data_char} is devoted to the statistical analysis of the available observations whose properties will be extended to full sky maps in section~\ref{fullsky}. 

\subsection{Differences with previous literature}\label{sec:oldlit}
Before entering in the details of the description of our method, we briefly describe some of the other 
techniques to model polarized synchrotron emission at low frequencies that have been carried out in the literature and their limitations. 
 \citet{Geil2011} and \citet{Jelic2008} both use the total intensity synchrotron emission $I$ as a template for polarized emission:
\begin{equation*}
Q(\hat{n},\nu_0)   =   p I(\hat{n},\nu_0) \cos(2\chi(\hat{n})) 
\end{equation*}\vspace{-0.5cm}
\begin{equation}
U(\hat{n},\nu_0)    =  p I(\hat{n},\nu_0) \sin(2\chi(\hat{n})) 
\end{equation}
where $p$ is the polarization fraction, function of the total intensity spectral index 
$\alpha_{\mathrm{syn}}$ \citep{LeRoux1961, Cortiglioni1995}:
\begin{equation}
p= \frac{3\alpha_{\mathrm{syn}}-3}{3\alpha_{\mathrm{syn}}-1}.
\end{equation}
In \citet{Jelic2008}, $\alpha_{\mathrm{syn}}$ is drawn from a 4D (3 spatial plus 1 frequency) Gaussian distribution 
while \citet{Geil2011} uses a running spectral index drawn from a Gaussian distribution. In \citet{Geil2011} the propagation through the ISM that leads to the Faraday modulation of the Stokes $Q$ and $U$ parameters is modelled using a single Faraday screen with $1 \lesssim \psi \lesssim 5~ \mathrm{rad}~\mathrm{m}^{-2}$ and an angular distribution with a small gradient over the sky-plane $\nabla \psi (x,y) \propto \hat{\mathbf{x}} +  \hat{\mathbf{y}}$. \citet{Jelic2008} uses two Faraday screens with zero mean, standard deviation 0.3 and a power law power spectrum with power law index arbitrarily set to $-2$. 

\citet{Alonso2014} and \citet{Shaw2015} use a different approach to the problem and simulate polarized emission directly in Faraday space, assuming its angular correlation function to be a power law with a correlation length $\zeta_{\psi}$. The Faraday depth $\psi$ is chosen to be a normally distributed variable around zero with variance $\sigma_{\psi}(\hat{n})$ determined from the full-sky Faraday rotation map of \citet{Oppermann2012}. Simulated maps are normalised to have a $20-30$\% average polarization fraction at high Galactic latitudes, according to the 23~GHz WMAP results \citep{Kogut2007}. 
\\

The aforementioned approaches have limitations in reproducing the characteristics of Galactic polarized emission below 200~MHz. Low frequency observations show that Galactic polarized emission is essentially ubiquitous at the Kelvin rms level 
both in selected, small sky patches \citep[e.g.][]{Bernardi2009,Bernardi2010,Iacobelli2013,Jelic2014} as well as in large scale surveys \citep[e.g.][]{Bernardi2013,Lenc2016}. Its spatial structure is often patchy and extends from arcmin to a few degree scales, although filamentary structures in the form of narrow, elongated ``canals" have been occasionally observed \citep{Jelic2015}. Its spatial structure is found to be statistically well described by a power spectrum in the usual multipole $\ell$ space \citep[e.g.][]{Bernardi2009,Jelic2014,Iacobelli2013}, i.e. $C_{\ell} \propto \ell^{\beta}$
with $\beta \sim -1.5$, flatter compared to measurements at cm-wavelengths \citep[e.g.][]{Carretti2005,Laporta2008}. 

One of the key features of the observed polarized emission at low frequencies that is not captured in the current simulation approaches is the almost complete lack of spatial correlation between total intensity and polarized emission due to a combination of observational effects: interferometric observations intrinsically filter out the large scale emission and are more sensitive to small scale thermal structure in the ISM \citep[e.g.][]{Wieringa1993,Gaensler2001,Bernardi2003}. Low frequency polarized emission is also structured along virtually any line of sight, showing emission peaks at different $\psi$ values \citep[e.g.][]{Schnitzeler2009,Bernardi2013,Lenc2016} - unlike the cm-wavelength regime. Such structures can be represented using the complex polarization $\tilde{P}$ as a function of Faraday depth $\psi$ described in section~\ref{theory}.

The spatial co-location of synchrotron emitting and Faraday rotating plasma - a common situation in the ISM - may originate multiple peaks in the Faraday spectrum $\tilde{P}(\psi)$ that can be separated at low frequencies due to the high $\psi$ resolution. Most of the Faraday peaks are observed at small $\psi$ values, consistent with a local (within a few hundreds pc) origin of the polarized emission \citep[][]{Haverkorn2004,Bernardi2013,Lenc2016}. Current simulation methods fall short, in general, to account for such Faraday structures. The use of the all-sky rotation measure (RM) map derived from extragalactic radio sources \citet{Alonso2014} is likely correct at GHz-frequencies but substantially overestimates the distribution of Faraday depths at low frequencies as it is integrated over the whole Galactic halo. \cite{Jelic2010} take into account various structures in Faraday depths, although they limit themselves to a few empirical models.

For our simulations, we will relax the assumption that total and polarized emission are spatially correlated and we use a realistic statistical representation of the angular and Faraday properties of the polarized emission driven by the available data.

\subsection{Simulation recipe}\label{simu_strat}

The goal of our simulations is to generate Stokes $Q$ and $U$ parameters at any sky direction $\hat{n}$ and frequency $\nu$\footnote{In this section we will interchangeably use $\nu$ and $\lambda^2$ to indicate the frequency dependence of the Stokes parameters as $\nu$ is the observer's variable and $\lambda^2$ is the proper variable to describe the properties in Faraday space.}. 

In the previous section we discussed how modelling $P(\hat{n}, \lambda^2)$ is difficult because of our limited knowledge of the detailed processes that occur in the ISM at low frequencies. We therefore decide to take advantage of the Fourier relationship that exists between $P (\lambda^2)$ and $\tilde{P} (\psi)$ (equation~\ref{RM1}) and generate $\tilde{P}(\hat{n}, \psi)$ maps whose statistics is constrained by current observations. Low frequency observations \citep[e.g.,][]{Bernardi2009,Jelic2014} justify the assumption that Stokes $\tilde{Q} (\hat{n}, \psi)$ ($\tilde{U} (\hat{n}, \psi)$) maps can be described as Gaussian distributed with a spatial structure of:
\begin{equation}\label{eq:A_alpha}
\langle \tilde{q}_{\ell m}(\psi) \tilde{q}^*_{\ell' m'}(\psi) \rangle = (2 \pi)^2 C_{\ell} (\psi) \delta_{\ell \ell'} \delta_{m m'}= (2 \pi)^2 A(\psi) \ell^{-\alpha(\psi)},
\end{equation}
where $\tilde{q}_{\ell m}$ are the coefficients of the spherical harmonics decomposition:
\begin{equation}
\tilde{Q}(\hat{n}, \psi)=\sum_{\ell m}\tilde{q}_{\ell m}(\psi)Y_{\ell m}(\hat{n})
\end{equation}
and $C_{\ell}(\psi)$ is the power spectrum.

The simulated maps $Q(\hat{n},\lambda^2)$ are then obtained through a Fourier transform of $\tilde{Q}(\hat{n}, \psi)$:
\begin{equation}\label{eq:Q_RM}
Q(\hat{n},\lambda^2)=\int \tilde{Q}(\hat{n}, \psi)e^{2 \pi i \lambda^2 \psi} d\psi.
\end{equation}
In the next section we explain how the power spectrum amplitude $A(\psi)$ and slope $\alpha(\psi)$ are constrained using available observations.
We note that the correlation structure along $\psi$ is taken automatically into account when imposing the normalisation  $A(\psi)$.

\subsection{Constraining simulation parameters} \label{data_char}

\begin{figure}
\includegraphics[width=\columnwidth]{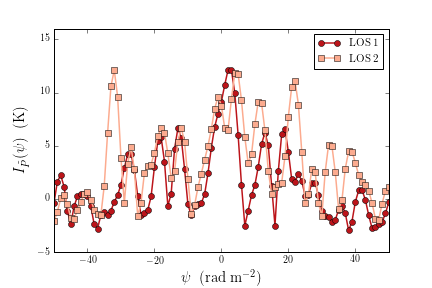}
\caption{Polarized intensity as a function of Faraday depth for two arbitrarily different lines of sight of the B13 data. The $\psi$-cube ranges from $\psi=-50$ to $\psi=+50~\mathrm{rad}~\mathrm{m}^{-2}$ with $\delta \psi=1$~rad~m$^{-2}$.}
\label{fig:LOS}
\end{figure}

In the previous section we showed that our simulations are specified by the power spectrum amplitude $A(\psi)$ and the slope $\alpha(\psi)$. As we are interested in large-scale simulations, we used the $2400$~square degree survey carried out with the Murchison Wide-field Array \citep[MWA,][]{Tingay2013} prototype at 189~MHz \citep[][, hereafter B13, the largest low frequency polarization survey available to date]{Bernardi2013} to constrain our model parameters. Their output is a $I_{\tilde{P}} (\hat{n}, \psi)$ cube that ranges from $\psi=-50$ to $\psi=+50~\mathrm{rad}~\mathrm{m}^{-2}$ with $\delta \psi=1$~rad~m$^{-2}$ and with a 15.6~arcmin angular resolution - although scales larger than $\sim 1^{\circ}$ are filtered out from the images. They have a 4.3~rad~m$^{-2}$ Faraday resolution and sample structures up to 1.5~rad~m$^{-2}$ - we refer the reader to the original paper for further details. Figure~\ref{fig:LOS} shows the polarized intensity $I_{\tilde P} (\psi)$ for two representative lines of sight.

\begin{figure}
\includegraphics[width=\columnwidth]{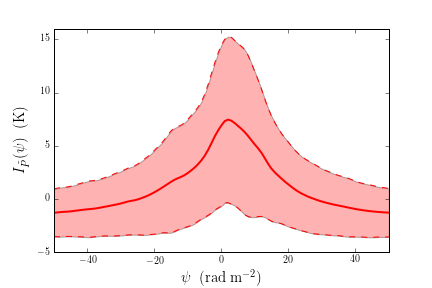}
\includegraphics[width=\columnwidth]{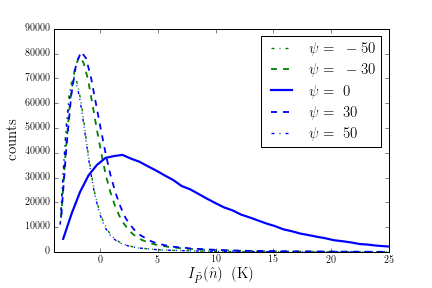}
\includegraphics[width=\columnwidth]{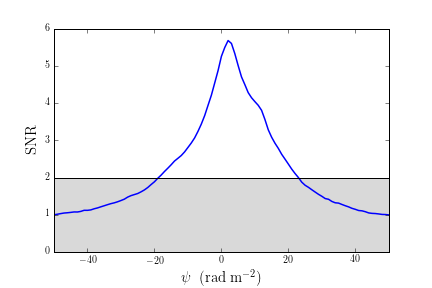}
\caption{Top: average polarized emission (solid line) and its $1\sigma$ standard deviation (shaded region) as a function of $\psi$ for the B13 data. The average peaks at $\approx 7~\mathrm{K}$ around $\psi \sim 0$~rad~m$^{-2}$ and flattens out at $|\psi| > 40$~rad~m$^{-2}$. Middle: the histogram of $I_{\tilde{P}}(\hat{n})$ for some selected $\psi$ values. 
Bottom: signal-to-noise ratio of the best fit Rayleigh parameter as a function of $\psi$ (see text for details). The grey region indicates the $2\sigma$ limit below which the polarized intensity slices are consistent with noise-like emission.}
\label{fig:global}
\end{figure}

The first step in order to evaluate [$A(\psi)$,$\alpha(\psi)$] is to investigate the behaviour of $I_{\tilde{P}}(\psi)$ once averaged over all the lines of sight. Figure~\ref{fig:global}, top panel, shows such an average as well as the standard deviation for each $\psi$ map. The curve clearly peaks around $\psi \sim 0$~rad~m$^{-2}$.  The middle panel of figure~\ref{fig:global}, instead, shows the histogram of $I_{\tilde{P}}(\hat{n})$ for some representative values of $\psi$. The histograms are similar for symmetric values of $\psi$ around zero. The signal dominates, as expected, around $\psi=0$~rad~m$^{-2}$. These histograms can be approximated by a Rayleigh distribution $R$, with a probability density function: 
\begin{equation}
R(x;\sigma_\psi)=\frac{x}{\sigma^2_\psi}e^{-x^2/2\sigma^2_\psi}, ~x\geq 0
\label{eq:ray}
\end{equation}
where  $\sigma_\psi$ is the scale parameter of the distribution\footnote{The maps from B13, whose histograms are shown in figure~\ref{fig:global}, middle panel, are noise-subtracted and therefore contains negative values. The Rayleigh distribution, on the contrary, is defined only for positive values.  Since we are only interested in the parameter $\sigma_\psi$, we added back the offset to perform the fit without this biasing the result.}. 
The compatibility of the data with a Rayleigh distribution justifies the choice of Gaussian distributed $\tilde{Q} (\hat{n}, \psi)$ and $\tilde{U} (\hat{n}, \psi)$ maps in equation~\ref{eq:A_alpha}: a Rayleigh distributed variable $R$ can be obtained using $R^2=X^2+Y^2$ , with $X,Y$ Gaussian distributed with null mean and equal variance. \\

We fitted a Rayleigh function to the distribution of the pixel values $I_{\tilde{P}}(\hat{n})$  for each $\psi$ slice of the B13 data cube. The best fit values ${\bar{\sigma}}_\psi$ for the parameter $\sigma_\psi$  are nearly constant for high $\psi$. We assumed this to be an empirical estimate of the noise level in the polarized intensity cube $I_{\tilde{P}}(\hat{n},\psi)$. In figure~\ref{fig:global}, bottom panel, we plot the signal-to-noise ratio ${\rm SNR} (\psi)$:
\begin{equation}
{\rm SNR} (\psi) = \frac{{\bar{\sigma}}_\psi}{\bar{\sigma}_{\mathrm{N}}}, \, {\rm with} \, \, \bar{\sigma}_{\mathrm{N}} = \frac{\bar{\sigma}_{-50} + \bar{\sigma}_{+50}}{2},
\label{eq:snr}
\end{equation}

where $\bar{\sigma}_{\mathrm{N}}$ is the noise level estimated using the more external slice of the data cube: $\psi=-50$ and $+50$. The ${\rm SNR} (\psi)$ curve qualitatively follows the polarized intensity $I_{\tilde{P}}(\psi)$ distribution, again suggesting that most of the sky emission is detected at small $\psi$ values. We estimate the noise in the polarized intensity data cube to be ${\tilde \sigma}_\psi = 1.5$~K and, in the following analysis of the spatial distribution of the polarized intensity, we will use only slices above the $2\sigma$ level, corresponding to $-18 < \psi < 23$~rad~m$^{-2}$. Also, to take into account the effect of the RMTF, we bin the data using $\delta \psi =3$~rad~m$^{-2}$. \\

The second step is to calculate the angular power spectrum $\hat{C}_{\ell}$ of the B13 data as a function of Faraday depth. We resampled the B13 images on the Healpix \citep{Gorski2004} grid with $N_{\mathrm{side}}=512$ and 
calculated the power spectrum using the routines \texttt{map2alm} and \texttt{alm2cl}, where the power spectrum of a generic function $f(\hat{n})$ is obtained with the standard estimator:
\begin{equation}
\hat{C}_{\ell}=\frac{1}{2\ell +1}\sum_m|\hat{a}_{\ell m}|^2,
\end{equation}
with 
\begin{equation}
\hat{a_{\ell m}}=\frac{4\pi}{N_{\mathrm{pix}}}\sum_{p=0}^{N_{\mathrm{pix}-1}}Y^{*}_{\ell m}(\hat{n}_p)f(\hat{n}_p).
\end{equation}

\begin{figure}
\includegraphics[width=\columnwidth]{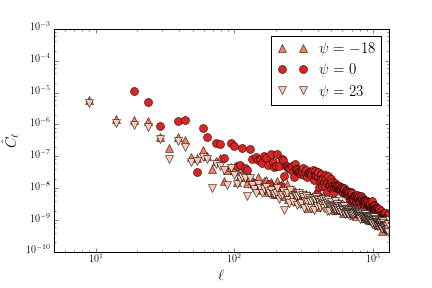}
\caption{Angular power spectrum $\hat{C}_{\ell}$ for three representative $\psi$ values, after deconvolution for incomplete sky coverage (see text). Values are in arbitrary units and errors are not included for the clarity of the figure.}
\label{fig:data_pl}
\end{figure}

\begin{figure}
\includegraphics[width=\columnwidth]{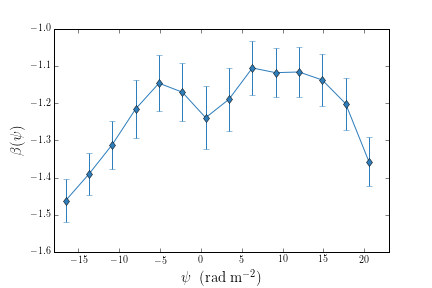}
\caption{Power spectrum slope $\beta(\psi)$ from the B13 data as a function of Faraday depth $\psi$. Errors bars are the sum of the uncertainties estimated by the MASTER algorithm and the pixel thermal noise \citep{Tegmark1997}. Only the range $50<\ell<700$ is considered in the fit.}
\label{fig:data_alpha}
\end{figure}

We correct for the effect of incomplete sky coverage using the MASTER algorithm \citep{Hivon2002}. Figure~\ref{fig:data_pl} shows an example of power spectra computed from the data. It fairly follows a power-law behaviour in $\ell$ at any observed $\psi$ value. 
Given the limited angular scales sample by the data, the power spectrum slope fits were limited to the $50 < \ell < 700$ range.  In the fit, we associated to the raw points calculated by MASTER an error at large scales due to cosmic variance and at small scales due to the thermal noise \citep{Tegmark1997}. We use for the r.m.s noise the estimated value $\bar{\sigma}_{\mathrm{N}}$ from equation~\ref{eq:snr}. 
We find slopes in the $-1.5 < \beta_\psi < -1.1$ range (figure~\ref{fig:data_alpha}), consistent with what was observed in much smaller sky patches \citep[e.g.,][]{Bernardi2009,Jelic2014}.

We note that the power spectrum slope tends to steepen at high $|\psi|$ values, although, given the error bars, it is consistent with an average value of $\beta = -1.3$ across the observed range.
In the next section we describe how we construct full-sky maps that preserve the observed statistics.

\subsection{Full-sky extrapolation}\label{fullsky}

\begin{figure}
\includegraphics[width=\columnwidth]{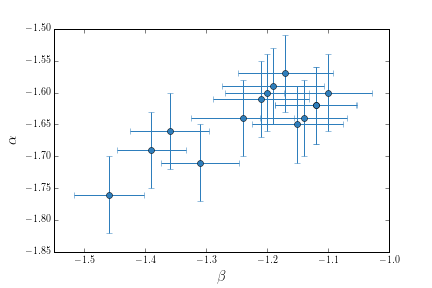}
\caption{The mapping between $\beta$, the power law index for the angular power spectrum $C_{\ell}^{\tilde{P}}=\ell^{\beta}$ fitted from polarization intensity MWA data and the power law index $\alpha$ used for simulating $\tilde{Q}$ and $\tilde{U}$. }
\label{fig:avsb}
\end{figure}

We now want to construct full-sky Stokes $Q$ and $U$ simulated maps whose statistics is consistent with what we derived from the polarized intensity maps in the previous section. In particular, we need to generate Stokes $\tilde{Q}$ and $\tilde{U}$ maps according to two independent Gaussian distributions (assumed to have the same mean and variance) that lead to a polarized intensity map with a spatial correlation that follows $C_{\ell}^{I_{\tilde{P}}}(\beta(\psi)) \propto \ell^{\beta(\psi)}$. This is not a trivial problem so we address it through a Monte Carlo approach, assuming also $\tilde{Q}$ and $\tilde{U}$ have a power law power spectrum. 
This allows us to determine the best value of $\alpha(\psi)$ (equation~\ref{eq:A_alpha}) that leads to the observed power spectrum slope $\beta(\psi)$. Each Monte Carlo realization includes the following steps:
\begin{enumerate}
\item 
for each $\beta_i$ value, corresponding to the $i$-th $\beta$ in figure~\ref{fig:data_alpha}, we scan a set of values $\{\alpha_{ij}\}_j$ carefully chosen around $\beta_i$ and generate, for every $\alpha_{ij}$ a $\tilde{Q}_{ij}$ and a $\tilde{U}_{ij}$ map with  $C_{\ell}^i=\ell^{\alpha_{ij}}$;
\item 
we calculate the polarized intensity maps $I_{\tilde{P}_{ij}} = \sqrt{\tilde{Q}_{ij}^2 + \tilde{U}_{ij}^2}$;
\item 
we fit a power law to the polarized intensity power spectrum $C_{\ell}^{I_{\tilde{P}}}(\alpha_{ij}) \propto\ell^{\hat{\beta} (\alpha_{ij})}$ to obtain the best fit slope $\hat{\beta}(\alpha_{ij})$ of the realization.
\end{enumerate}
The above procedure is repeated for $N=10$ realizations for every value of $\alpha_{ij}$.  
For each given $\beta_i$, we construct the estimator:
\begin{equation}
\chi_j^2(\alpha_{ij} | \beta_i) \equiv  \sum_{\ell} \frac{(\hat{C}_{\ell}^{I_{\tilde{P}}}(\alpha_{ij})-C_{\ell}^{I_{\tilde{P}}}(\beta_i))^2}{\sigma_{C_{\ell}^{I_{\tilde{P}}}(\beta_i)}}  
\end{equation}
as a function of $\alpha_{ij}$ and for the $N$ different realizations. As an estimation of $\sigma_{C_{\ell}(\beta_i)}$ we consider only the cosmic variance contribution, i.e. $\sigma_{C_{\ell}^{I_{\tilde{P}}}(\beta_i)}=\sqrt{\frac{2}{2\ell+1}}C_{\ell}^{I_{\tilde{P}}}(\beta_i)$. 
We then select, for every $\beta_i$, the value of $\alpha_{ij}$ that minimizes the mean over the $N$ realizations of the value of $\chi_j^2(\alpha_{ij} | \beta_i)$. This allows us to map the values of $\beta$, i.e. the spectral index of the $I_{\tilde{P}}$, into values of $\alpha$, i.e. the corresponding spectral index for the $\tilde{Q}$ and $\tilde{U}$ maps,  as shown in figure~\ref{fig:avsb}.

At this point, the desired maps can be obtained from the coefficients of the spherical harmonic expansion:

\begin{equation}\label{almQU}
a_{\ell m}^{\tilde{Q},\tilde{U}}=\sqrt{\frac{C_{\ell}}{2}}N(0,1)+i \sqrt{\frac{C_{\ell}}{2}}N(0,1).
\end{equation}
As a function of $\psi$, the maps simulated in this way have the desired angular power spectrum and the pixel values follow, by construction, a Gaussian distribution with zero mean and variance $4\pi\sum_{\ell}C_{\ell}$. We renormalised them to the variance ${\bar{\sigma}}_{\psi}$ in order to produce the values observed in the B13 data (section~\ref{data_char}).

\begin{figure}
\includegraphics[width=\columnwidth]{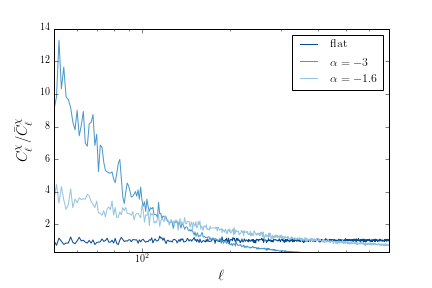}
\caption{The angular power spectrum $C_{\ell}^{\chi}$  for the phase $\chi$ of $P$, normalised with respect to its mean value $\bar{C}_{\ell}^{\chi}$ in the range  $50< \ell <700$,  for a steep power law value of $-3$ (in blue) in comparison with a typical value of $-1.6$ found in this work (light blue). The case of a flat power spectrum for $\chi$ is also shown as reference (dark blue).}
\label{fig:phase}
\end{figure}

We can calculate the polarisation angle $\chi$ using equation~\ref{eq:phase}, pixel by pixel.
By construction, with generic simulated Stokes $\tilde{Q}$ and $\tilde{U}$ maps that follow a Gaussian distribution, one obtains a uniform distribution in the polarization angle. However, when one imposes a spatial structure with a steep power law to $\tilde{Q}$ and $\tilde{U}$, the spatial behaviour of the phase deviates from just the flat power spectrum. In figure~\ref{fig:phase} we show that, for the power law values considered in this analysis, the polarization angle spatial structure is still reasonably flat and thus uninformative. Note that the knowledge of the absolute polarization angle is not necessary for the purpose of our simulations, but only if we were interested to simulate the orientation of the magnetic field. 

Since we extract constraints from polarized intensity without any information on the phase, we can always argue that this information can be added from external data. We can easily include a {\it phase map} $\chi^{\prime}$ constructing a new $\tilde{P}^{\prime}=I_{\tilde{P}}e^{2i\chi^{\prime}}=\tilde{Q}^{\prime}+i\tilde{U}^{\prime}$. The resulting $\tilde{Q}^{\prime}$ and $\tilde{U}^{\prime}$ maps will not be the original one of the procedure, but the total polarized intensity structure will still hold and will show the desired power spectrum. 

\section{Results and discussion}\label{sec:results}
In section~\ref{sec:sim} we described the details of our simulations to the 
production of full-sky $\tilde{Q}$ and $\tilde{U}$ cubes that cover the $-18 < \psi < 23 $ range in steps of $\delta \psi \simeq 3~\mathrm{rad}~\mathrm{m}^{-2}$. In this this section, we verify the consistency of our results with the data  and discuss qualitatively their implications in term of contamination of the $21$-cm signal.

\begin{figure}
\includegraphics[width=\columnwidth]{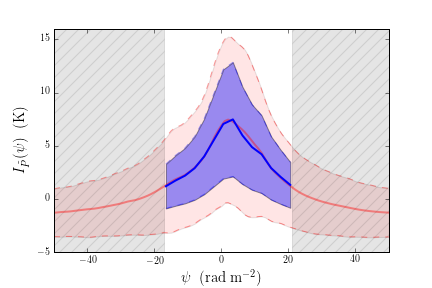}
\caption{Mean (blue solid line) and $1\sigma$ fluctuations (blue shaded region) as a function of $\psi$ for the full-sky simulated maps of polarized intensity. The mean agrees with the average computed from the B13 data (red line, same as figure~\ref{fig:global}). We recall we only use and then simulate the interval $-18 < \psi < 23 ~\mathrm{rad}~\mathrm{m}^{-2}$.}
\label{fig:simu_test}
\end{figure}

\begin{figure}
\includegraphics[width=\columnwidth]{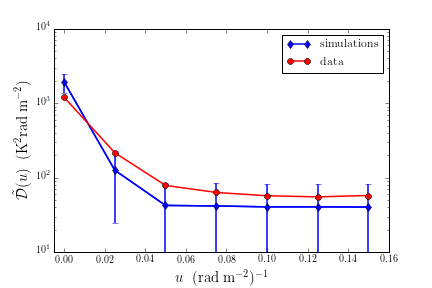}
\caption{Comparison between the correlation in the data and in the simulations. $\tilde{\mathcal{D}}(u)$ is calculated as the mean value of the power spectrum over every line of sight in the Faraday depth cube as a function of $\psi$, with bins of $\delta \psi \simeq 3$.}
\label{fig:simu_corr}
\end{figure}

Figure~\ref{fig:simu_test} shows a consistency check against the input polarized intensity distribution $I_{\tilde{P}}$ whereas figure~\ref{fig:simu_corr} shows a comparison between the data and the simulation power spectra computed as: 
\begin{equation}
\langle \hat{I}_{\tilde{P}}(u)  \hat{I}^{*}_{\tilde{P}}(u) \rangle = \tilde{\mathcal{D}} (u) \Delta \psi / (\delta\psi)^2
\end{equation}
where $\hat{I}_{\tilde{P}}(u)$ is the FFT of $I_{\tilde{P}}(\psi)$, $\langle \cdot \rangle$ indicates the average over the different lines of sight, $\delta \psi$ is the bin width in $\psi$ and $\Delta \psi$ is the range in $\psi$ covered by the simulations.  Both tests confirm that our simulated maps follow the data statistical properties.

\begin{figure}
\includegraphics[width=0.5\textwidth]{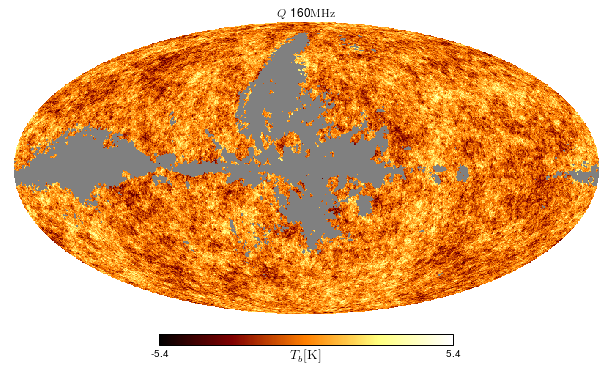}
\includegraphics[width=0.5\textwidth]{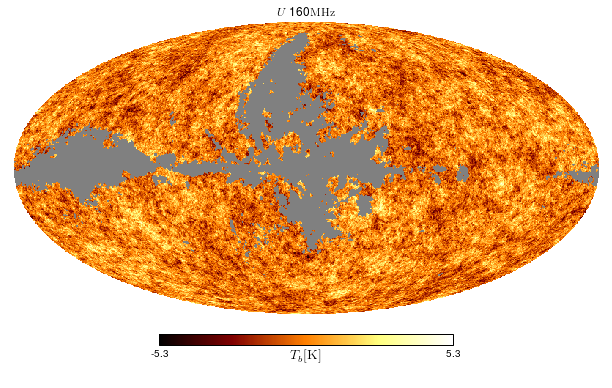}
\includegraphics[width=0.5\textwidth]{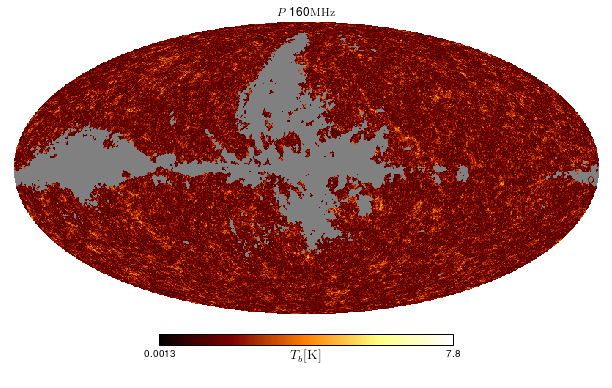}
\caption{Examples of simulated polarized $Q$, $U$ and polarized intensity synchrotron maps at $160$~MHz. The mask is calculated from WMAP 23~GHz retaining~$f_{\mathrm{sky}}=90\%$ in $Q$ and $U$.}
\label{fig:simu_189}
\end{figure}

The simulated cube is converted to frequency space through equation~\ref{eq:Q_RM}. The simulated maps at 160~MHz are shown in figure~\ref{fig:simu_189} as an example. We blanked out the regions of brightest emission in the 23~GHz WMAP polarized intensity map where real observations may show a significant deviation from the statistics used in our simulations.
However, as discussed in section~\ref{intro}, the low global level of polarization measured by MWA in comparison with the expectation obtained from a standard power-law extrapolation from higher frequencies, reveals that the emission has a local origin. Indeed, the polarization horizon \citep{Landecker2001}, i.e. the maximal distance from the observer beyond which the emission is depolarized, seems to diminish consistently going to the low frequencies of interest (e.g. \citet{Bernardi2003,Brouw1976}). Both \citet{Bernardi2013} and \citet{Lenc2016}, from their measured values of $\psi$ and using equation~\ref{eq:psi} with simplistic assumptions on the thermal electron density and the magnetic field strength, found that the polarization horizon is not farther than $\sim120$ pc. \citet{Lenc2016} refines this estimation using $\psi$ measures from pulsar, bringing the polarization horizon down to  $\sim 50$ pc.  These values implies that the structure of the Galactic plane\footnote{We remind that the Galactic center is $\sim 8$~Kpc distant from us.} in polarization, should be very difficult to see at low frequencies.\\

\begin{figure}
\includegraphics[width=\columnwidth]{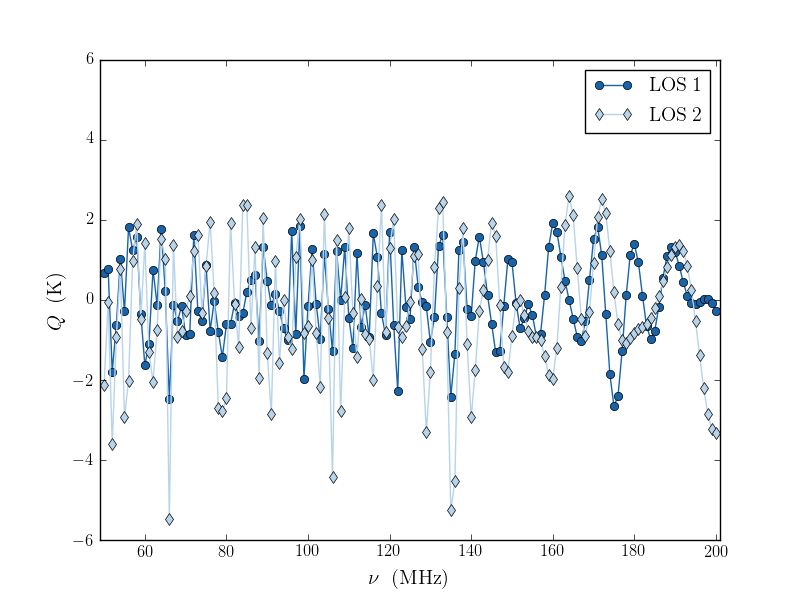}
\caption{Stokes $Q(\nu)$ as a function of frequency $\nu$ for two randomly chosen lines of sight.}
\label{fig:Tbvsfreq_QU}
\end{figure}

\begin{figure}
		\includegraphics[width=\columnwidth]{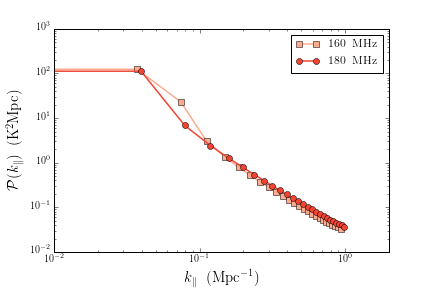}
    \caption{The power spectrum $\mathcal{P}(k_{\parallel})$ of the final simulated $Q(\hat{n},\nu)$ maps,  for a $10$~MHz bandwidth centred around $160$~MHz (light red squares) and $180$~MHz (red circles), averaged over all lines of sight. This corresponds to a smoothing scale in the perpendicular direction of about $k_\perp=0.055~\mathrm{Mpc}^{-1}$}
    \label{fig:radial_ps2}
\end{figure}

Our simulations allow us to quantify the expected frequency coherence of the Galactic polarized foreground: the Stokes $Q$ parameter for two arbitrary lines of sight is displayed in figure~\ref{fig:Tbvsfreq_QU}, where we can see that its sign changes on scales of a few MHz, not much different than the scale of coherence of the 21~cm signal.
Figure~\ref{fig:radial_ps2} offers a view of the coherence scale directly in $k_{\parallel}$ space. We use simulations of a $10$~MHz bandwidth  (where no significant cosmological evolution is expected to happen) centred around $160$ and $180$~MHz refining the frequency resolution to $0.1$~MHz. We calculated the power spectrum along the line of sight as:
\begin{equation}
\langle \hat{Q}(k_{\parallel})   \hat{Q}^*(k_{\parallel}) \rangle = \mathcal{P}(k_{\parallel}) \Delta r /(\delta r)^2
\end{equation}
where $\hat{Q}(k_{\parallel})$  is the FFT of $Q(\nu)$ re-sampled to be equispaced in the comoving distance $r(z)$. $\langle \cdot \rangle$ is again the average over all the lines of sight. $\Delta r$ is the interval in Mpc corresponding to the frequency range and  $\delta \nu$ is converted to cosmological distance $\delta r$ using:
\begin{equation}
\delta r = c H^{-1}(z) \frac{(1+z)^2}{\nu_{21}} \delta \nu,
\end{equation}
where $r(z)$ and $H(z)$ are obtained assuming {\it Planck} 2015 cosmology \citep{PlanckXIII} and $\nu_{21}=1420$~MHz is the rest frequency of the $21$~cm line. Note that since we are using a Healpix pixelization with $N_{\mathrm{side}}=512$, the power spectrum of figure~\ref{fig:radial_ps2} corresponds to a smoothing scale in the perpendicular direction of about $k_\perp=0.055~\mathrm{Mpc}^{-1}$, where $k_\perp=1/ (r(z)\Delta \theta)$ with $\Delta \theta$ the pixel size.

Power spectra at both frequencies are fairly similar, as expected, and both show a fast decline beyond $k_{\parallel} \sim 0.05$~Mpc$^{-1}$. As there is a linear relationship between the Faraday depth $\psi$ and $k_{\parallel}$ \citep{Pen2009,Moore2017,Nunhokee2017}, the decrease in power at high $k_{\parallel}$ is consistent with the fact that the simulated maps have most of the power at small $\psi$.

\section{Conclusions}\label{sec:conclusions}
In this paper we presented full-sky simulations of the Galactic polarized synchrotron emission in the $50-200$~MHz frequency range, relevant for 21~cm observations from the EoR. Unlike previous simulation methods, we did not use total intensity data as a proxy since there is a lack of correlation between total and polarized diffuse emission at low frequencies. We derived, instead, the statistical properties of large-scale observations of diffuse polarized emission at low frequencies \citep{Bernardi2013} and used them to simulate polarization maps directly in Faraday space that are then Fourier transformed to frequency space \citep[][]{Burn1966,Brentjens2005}. The simulated Stokes $Q$ and $U$ maps spanning the $50-200$~MHz interval with $1$~MHz resolution, are publicly available at \href{https://drive.google.com/drive/folders/0B0ACYlYtsnNISEhoZjVFaVJfV28?usp=sharing}{UWC - Center for Radio Cosmology Google Drive Directory}.

The polarized power spectra derived from simulations shows a steep declining power at $k_\parallel > 0.05$~Mpc$^{-1}$, consistent with the fact that polarized emission is mostly located at small Faraday depth values. At a qualitative level, this result is consistent with the simulations in \cite{Nunhokee2017} and, therefore, leave open the possibility that there is a power spectrum region where the polarization leakage contamination may just be below the 21~cm signal, mitigating the requirements for its modelling and subtraction.
In our case, however, the simulations predict the power spectrum amplitude and the shape more realistically than in previous literature and represent, therefore, better inputs for instrumental simulations. 

It is worth underlying that our simulation extrapolation to $50$~MHz has been based on the statistics from the $189$~MHz data and, therefore, did not account for possible depolarization effects occurring at lower frequencies or for the spectral dependence of the polarized emission.
As pointed out earlier, however, both these effects are largely unknown and our choice of extrapolating the properties of the $189$~MHz Faraday space to lower frequencies is, therefore, somewhat conservative. As data at lower frequencies become available, they can be directly included in our simulation framework.

Future work will focus on using our results in realistic simulation pipelines of experiments such as the Hydrogen Epoch of Reionization Array \citep{DeBoer2017} or the upcoming Square Kilometre Array \citep{Koopmans2015}. This will allow to predict the actual contamination level for different leakage terms and test foreground separation methods.

\section*{Acknowledgements}
MS and MGS are supported by the South African Square Kilometre Array Project and National Research Foundation.
GB acknowledges support from the Royal Society and the Newton Fund under grant NA150184. This work is based on research supported in part by the National Research Foundation of South Africa (grant No. 103424).




\bibliographystyle{mnras}
\bibliography{pol_synch} 

\begin{thebibliography}{}
\makeatletter
\relax
\def\mn@urlcharsother{\let\do\@makeother \do\$\do\&\do\#\do\^\do\_\do\%\do\~}
\def\mn@doi{\begingroup\mn@urlcharsother \@ifnextchar [ {\mn@doi@}
  {\mn@doi@[]}}
\def\mn@doi@[#1]#2{\def\@tempa{#1}\ifx\@tempa\@empty \href
  {http://dx.doi.org/#2} {doi:#2}\else \href {http://dx.doi.org/#2} {#1}\fi
  \endgroup}
\def\mn@eprint#1#2{\mn@eprint@#1:#2::\@nil}
\def\mn@eprint@arXiv#1{\href {http://arxiv.org/abs/#1} {{\tt arXiv:#1}}}
\def\mn@eprint@dblp#1{\href {http://dblp.uni-trier.de/rec/bibtex/#1.xml}
  {dblp:#1}}
\def\mn@eprint@#1:#2:#3:#4\@nil{\def\@tempa {#1}\def\@tempb {#2}\def\@tempc
  {#3}\ifx \@tempc \@empty \let \@tempc \@tempb \let \@tempb \@tempa \fi \ifx
  \@tempb \@empty \def\@tempb {arXiv}\fi \@ifundefined
  {mn@eprint@\@tempb}{\@tempb:\@tempc}{\expandafter \expandafter \csname
  mn@eprint@\@tempb\endcsname \expandafter{\@tempc}}}

\bibitem[\protect\citeauthoryear{{Ali} et~al.,}{{Ali} et~al.}{2015}]{Ali2015}
{Ali} Z.~S.,  et~al., 2015, \mn@doi [\apj] {10.1088/0004-637X/809/1/61}, \href
  {http://adsabs.harvard.edu/abs/2015ApJ...809...61A} {809, 61}

\bibitem[\protect\citeauthoryear{Alonso, Ferreira  \& Santos}{Alonso
  et~al.}{2014}]{Alonso2014}
Alonso D.,  Ferreira P.~G.,   Santos M.~G.,  2014, \mn@doi [\mnras]
  {10.1093/mnras/stu1666}, 444, 3183

\bibitem[\protect\citeauthoryear{{Asad} et~al.,}{{Asad}
  et~al.}{2015}]{Asad2015}
{Asad} K.~M.~B.,  et~al., 2015, \mn@doi [\mnras] {10.1093/mnras/stv1107}, \href
  {http://adsabs.harvard.edu/abs/2015MNRAS.451.3709A} {451, 3709}

\bibitem[\protect\citeauthoryear{{Asad} et~al.,}{{Asad}
  et~al.}{2016}]{Asad2016}
{Asad} K.~M.~B.,  et~al., 2016, \mn@doi [\mnras] {10.1093/mnras/stw1863}, \href
  {http://adsabs.harvard.edu/abs/2016MNRAS.462.4482A} {462, 4482}

\bibitem[\protect\citeauthoryear{{Beardsley} et~al.,}{{Beardsley}
  et~al.}{2016}]{Beardsley2016}
{Beardsley} A.~P.,  et~al., 2016, \mn@doi [\apj] {10.3847/1538-4357/833/1/102},
  \href {http://adsabs.harvard.edu/abs/2016ApJ...833..102B} {833, 102}

\bibitem[\protect\citeauthoryear{{Bernardi}, {Carretti}, {Cortiglioni},
  {Sault}, {Kesteven}  \& {Poppi}}{{Bernardi} et~al.}{2003}]{Bernardi2003}
{Bernardi} G.,  {Carretti} E.,  {Cortiglioni} S.,  {Sault} R.~J.,  {Kesteven}
  M.~J.,   {Poppi} S.,  2003, \mn@doi [\apjl] {10.1086/378398}, \href
  {http://adsabs.harvard.edu/abs/2003ApJ...594L...5B} {594, L5}

\bibitem[\protect\citeauthoryear{{Bernardi} et~al.,}{{Bernardi}
  et~al.}{2009}]{Bernardi2009}
{Bernardi} G.,  et~al., 2009, \mn@doi [\aap] {10.1051/0004-6361/200911627},
  \href {http://adsabs.harvard.edu/abs/2009A%26A...500..965B} {500, 965}

\bibitem[\protect\citeauthoryear{{Bernardi} et~al.,}{{Bernardi}
  et~al.}{2010}]{Bernardi2010}
{Bernardi} G.,  et~al., 2010, \mn@doi [\aap] {10.1051/0004-6361/200913420},
  \href {http://adsabs.harvard.edu/abs/2010A%26A...522A..67B} {522, A67}

\bibitem[\protect\citeauthoryear{{Bernardi} et~al.,}{{Bernardi}
  et~al.}{2013}]{Bernardi2013}
{Bernardi} G.,  et~al., 2013, \mn@doi [\apj] {10.1088/0004-637X/771/2/105},
  \href {http://adsabs.harvard.edu/abs/2013ApJ...771..105B} {771, 105}

\bibitem[\protect\citeauthoryear{{Bernardi} et~al.,}{{Bernardi}
  et~al.}{2016}]{Bernardi2016}
{Bernardi} G.,  et~al., 2016, \mn@doi [\mnras] {10.1093/mnras/stw1499}, \href
  {http://adsabs.harvard.edu/abs/2016MNRAS.461.2847B} {461, 2847}

\bibitem[\protect\citeauthoryear{Brentjens \& de Bruyn}{Brentjens \&
  de~Bruyn}{2005}]{Brentjens2005}
Brentjens M.~a.,  de Bruyn a.~G.,  2005, \mn@doi [\aap]
  {10.1051/0004-6361:20052990}, 441, 1217

\bibitem[\protect\citeauthoryear{{Brouw} \& {Spoelstra}}{{Brouw} \&
  {Spoelstra}}{1976}]{Brouw1976}
{Brouw} W.~N.,  {Spoelstra} T.~A.~T.,  1976, \aaps, \href
  {http://adsabs.harvard.edu/abs/1976A%26AS...26..129B} {26, 129}

\bibitem[\protect\citeauthoryear{Burn}{Burn}{1966}]{Burn1966}
Burn B.~J.,  1966, \mn@doi [\mnras] {10.1093/mnras/133.1.67}, 133, 67

\bibitem[\protect\citeauthoryear{{Carretti}, {Bernardi}, {Sault}, {Cortiglioni}
   \& {Poppi}}{{Carretti} et~al.}{2005}]{Carretti2005}
{Carretti} E.,  {Bernardi} G.,  {Sault} R.~J.,  {Cortiglioni} S.,   {Poppi} S.,
   2005, \mn@doi [\mnras] {10.1111/j.1365-2966.2005.08761.x}, \href
  {http://adsabs.harvard.edu/abs/2005MNRAS.358....1C} {358, 1}

\bibitem[\protect\citeauthoryear{{Chapman}, {Zaroubi}, {Abdalla}, {Dulwich},
  {Jeli{\'c}}  \& {Mort}}{{Chapman} et~al.}{2016}]{Chapman2016}
{Chapman} E.,  {Zaroubi} S.,  {Abdalla} F.~B.,  {Dulwich} F.,  {Jeli{\'c}} V.,
   {Mort} B.,  2016, \mn@doi [\mnras] {10.1093/mnras/stw161}, \href
  {http://adsabs.harvard.edu/abs/2016MNRAS.458.2928C} {458, 2928}

\bibitem[\protect\citeauthoryear{{Cortiglioni} \& {Spoelstra}}{{Cortiglioni} \&
  {Spoelstra}}{1995}]{Cortiglioni1995}
{Cortiglioni} S.,  {Spoelstra} T.~A.~T.,  1995, \aap, \href
  {http://adsabs.harvard.edu/abs/1995A%26A...302....1C} {302, 1}

\bibitem[\protect\citeauthoryear{{DeBoer} et~al.,}{{DeBoer}
  et~al.}{2017}]{DeBoer2017}
{DeBoer} D.~R.,  et~al., 2017, \mn@doi [\pasp]
  {10.1088/1538-3873/129/974/045001}, \href
  {http://adsabs.harvard.edu/abs/2017PASP..129d5001D} {129, 045001}

\bibitem[\protect\citeauthoryear{{Dillon} et~al.,}{{Dillon}
  et~al.}{2014}]{Dillon2014}
{Dillon} J.~S.,  et~al., 2014, \mn@doi [\prd] {10.1103/PhysRevD.89.023002},
  \href {http://adsabs.harvard.edu/abs/2014PhRvD..89b3002D} {89, 023002}

\bibitem[\protect\citeauthoryear{{Dillon} et~al.,}{{Dillon}
  et~al.}{2015}]{Dillon2015}
{Dillon} J.~S.,  et~al., 2015, \mn@doi [\prd] {10.1103/PhysRevD.91.123011},
  \href {http://adsabs.harvard.edu/abs/2015PhRvD..91l3011D} {91, 123011}

\bibitem[\protect\citeauthoryear{{Ewall-Wice} et~al.,}{{Ewall-Wice}
  et~al.}{2016}]{Ewall-Wice2016}
{Ewall-Wice} A.,  et~al., 2016, \mn@doi [\mnras] {10.1093/mnras/stw1022}, \href
  {http://adsabs.harvard.edu/abs/2016MNRAS.460.4320E} {460, 4320}

\bibitem[\protect\citeauthoryear{{Furlanetto}}{{Furlanetto}}{2016}]{Furlanetto2016}
{Furlanetto} S.~R.,  2016, in {Mesinger} A.,  ed.,  Astrophysics and Space
  Science Library Vol. 423, Understanding the Epoch of Cosmic Reionization:
  Challenges and Progress. p.~247 (\mn@eprint {arXiv} {1511.01131}),
  \mn@doi{10.1007/978-3-319-21957-8_9}

\bibitem[\protect\citeauthoryear{{Gaensler}, {Dickey}, {McClure-Griffiths},
  {Green}, {Wieringa}  \& {Haynes}}{{Gaensler} et~al.}{2001}]{Gaensler2001}
{Gaensler} B.~M.,  {Dickey} J.~M.,  {McClure-Griffiths} N.~M.,  {Green} A.~J.,
  {Wieringa} M.~H.,   {Haynes} R.~F.,  2001, \mn@doi [\apj] {10.1086/319468},
  \href {http://adsabs.harvard.edu/abs/2001ApJ...549..959G} {549, 959}

\bibitem[\protect\citeauthoryear{{Geil}, {Gaensler}  \& {Wyithe}}{{Geil}
  et~al.}{2011}]{Geil2011}
{Geil} P.~M.,  {Gaensler} B.~M.,   {Wyithe} J.~S.~B.,  2011, \mn@doi [\mnras]
  {10.1111/j.1365-2966.2011.19509.x}, \href
  {http://adsabs.harvard.edu/abs/2011MNRAS.418..516G} {418, 516}

\bibitem[\protect\citeauthoryear{Gorski, Hivon, Banday, Wandelt, Hansen,
  Reinecke  \& Bartelman}{Gorski et~al.}{2004}]{Gorski2004}
Gorski K.~M.,  Hivon E.,  Banday A.~J.,  Wandelt B.~D.,  Hansen F.~K.,
  Reinecke M.,   Bartelman M.,  2004, \mn@doi [\apj] {10.1086/427976}, 622, 759

\bibitem[\protect\citeauthoryear{{Haverkorn}, {Katgert}  \& {de
  Bruyn}}{{Haverkorn} et~al.}{2004}]{Haverkorn2004}
{Haverkorn} M.,  {Katgert} P.,   {de Bruyn} A.~G.,  2004, \mn@doi [\aap]
  {10.1051/0004-6361:200400051}, \href
  {http://adsabs.harvard.edu/abs/2004A%26A...427..549H} {427, 549}

\bibitem[\protect\citeauthoryear{{Heald}}{{Heald}}{2009}]{Heald2009b}
{Heald} G.,  2009, in {Strassmeier} K.~G.,  {Kosovichev} A.~G.,   {Beckman}
  J.~E.,  eds,  IAU Symposium Vol. 259, Cosmic Magnetic Fields: From Planets,
  to Stars and Galaxies. pp 591--602, \mn@doi{10.1017/S1743921309031421}

\bibitem[\protect\citeauthoryear{Hivon, G{\'{o}}rski, Netterfield, Crill,
  Prunet  \& Hansen}{Hivon et~al.}{2002}]{Hivon2002}
Hivon E.,  G{\'{o}}rski K.~M.,  Netterfield C.~B.,  Crill B.~P.,  Prunet S.,
  Hansen F.,  2002, \mn@doi [\apj] {10.1086/338126}, 567, 2

\bibitem[\protect\citeauthoryear{{Iacobelli} et~al.,}{{Iacobelli}
  et~al.}{2013}]{Iacobelli2013}
{Iacobelli} M.,  et~al., 2013, \mn@doi [\aap] {10.1051/0004-6361/201322013},
  \href {http://adsabs.harvard.edu/abs/2013A%26A...558A..72I} {558, A72}

\bibitem[\protect\citeauthoryear{{Jacobs} et~al.,}{{Jacobs}
  et~al.}{2015}]{Jacobs2015}
{Jacobs} D.~C.,  et~al., 2015, \mn@doi [\apj] {10.1088/0004-637X/801/1/51},
  \href {http://adsabs.harvard.edu/abs/2015ApJ...801...51J} {801, 51}

\bibitem[\protect\citeauthoryear{{Jeli{\'c}} et~al.,}{{Jeli{\'c}}
  et~al.}{2008}]{Jelic2008}
{Jeli{\'c}} V.,  et~al., 2008, \mn@doi [\mnras]
  {10.1111/j.1365-2966.2008.13634.x}, \href
  {http://adsabs.harvard.edu/abs/2008MNRAS.389.1319J} {389, 1319}

\bibitem[\protect\citeauthoryear{Jeli{\'{c}}, Zaroubi, Labropoulos, Bernardi,
  {De Bruyn}  \& Koopmans}{Jeli{\'{c}} et~al.}{2010}]{Jelic2010}
Jeli{\'{c}} V.,  Zaroubi S.,  Labropoulos P.,  Bernardi G.,  {De Bruyn} A.~G.,
   Koopmans L. V.~E.,  2010, \mn@doi [\mnras]
  {10.1111/j.1365-2966.2010.17407.x}, 409, 1647

\bibitem[\protect\citeauthoryear{{Jeli{\'c}} et~al.,}{{Jeli{\'c}}
  et~al.}{2014}]{Jelic2014}
{Jeli{\'c}} V.,  et~al., 2014, \mn@doi [\aap] {10.1051/0004-6361/201423998},
  \href {http://adsabs.harvard.edu/abs/2014A%26A...568A.101J} {568, A101}

\bibitem[\protect\citeauthoryear{{Jeli{\'c}} et~al.,}{{Jeli{\'c}}
  et~al.}{2015}]{Jelic2015}
{Jeli{\'c}} V.,  et~al., 2015, \mn@doi [\aap] {10.1051/0004-6361/201526638},
  \href {http://adsabs.harvard.edu/abs/2015A%26A...583A.137J} {583, A137}

\bibitem[\protect\citeauthoryear{{Kogut} et~al.,}{{Kogut}
  et~al.}{2007}]{Kogut2007}
{Kogut} A.,  et~al., 2007, \mn@doi [\apj] {10.1086/519754}, \href
  {http://adsabs.harvard.edu/abs/2007ApJ...665..355K} {665, 355}

\bibitem[\protect\citeauthoryear{{Kohn} et~al.,}{{Kohn}
  et~al.}{2016}]{Kohn2016}
{Kohn} S.~A.,  et~al., 2016, \mn@doi [\apj] {10.3847/0004-637X/823/2/88}, \href
  {http://adsabs.harvard.edu/abs/2016ApJ...823...88K} {823, 88}

\bibitem[\protect\citeauthoryear{{Koopmans} et~al.,}{{Koopmans}
  et~al.}{2015}]{Koopmans2015}
{Koopmans} L.,  et~al., 2015, Advancing Astrophysics with the Square Kilometre
  Array (AASKA14), \href {http://adsabs.harvard.edu/abs/2015aska.confE...1K}
  {p.~1}

\bibitem[\protect\citeauthoryear{{La Porta}, {Burigana}, {Reich}  \&
  {Reich}}{{La Porta} et~al.}{2008}]{Laporta2008}
{La Porta} L.,  {Burigana} C.,  {Reich} W.,   {Reich} P.,  2008, \mn@doi [\aap]
  {10.1051/0004-6361:20078435}, \href
  {http://adsabs.harvard.edu/abs/2008A%26A...479..641L} {479, 641}

\bibitem[\protect\citeauthoryear{{Landecker}, {Uyan{\i}ker}  \&
  {Kothes}}{{Landecker} et~al.}{2001}]{Landecker2001}
{Landecker} T.~L.,  {Uyan{\i}ker} B.,   {Kothes} R.,  2001, in American
  Astronomical Society Meeting Abstracts. p.~1390

\bibitem[\protect\citeauthoryear{{Le Roux}}{{Le Roux}}{1961}]{LeRoux1961}
{Le Roux} E.,  1961, Annales d'Astrophysique, \href
  {http://adsabs.harvard.edu/abs/1961AnAp...24...71L} {24, 71}

\bibitem[\protect\citeauthoryear{{Lenc} et~al.,}{{Lenc}
  et~al.}{2016}]{Lenc2016}
{Lenc} E.,  et~al., 2016, \mn@doi [\apj] {10.3847/0004-637X/830/1/38}, \href
  {http://adsabs.harvard.edu/abs/2016ApJ...830...38L} {830, 38}

\bibitem[\protect\citeauthoryear{{Liu}, {Parsons}  \& {Trott}}{{Liu}
  et~al.}{2014a}]{Liu2014a}
{Liu} A.,  {Parsons} A.~R.,   {Trott} C.~M.,  2014a, \mn@doi [\prd]
  {10.1103/PhysRevD.90.023018}, \href
  {http://adsabs.harvard.edu/abs/2014PhRvD..90b3018L} {90, 023018}

\bibitem[\protect\citeauthoryear{{Liu}, {Parsons}  \& {Trott}}{{Liu}
  et~al.}{2014b}]{Liu2014b}
{Liu} A.,  {Parsons} A.~R.,   {Trott} C.~M.,  2014b, \mn@doi [\prd]
  {10.1103/PhysRevD.90.023019}, \href
  {http://adsabs.harvard.edu/abs/2014PhRvD..90b3019L} {90, 023019}

\bibitem[\protect\citeauthoryear{{McQuinn}}{{McQuinn}}{2016}]{McQuinn2016}
{McQuinn} M.,  2016, {Deciphering the Cosmic Dawn with Lyman-alpha in Emission
  and Absorption}, NASA ATP Proposal

\bibitem[\protect\citeauthoryear{{Monsalve}, {Rogers}, {Bowman}  \&
  {Mozdzen}}{{Monsalve} et~al.}{2017}]{Monsalve2017}
{Monsalve} R.~A.,  {Rogers} A.~E.~E.,  {Bowman} J.~D.,   {Mozdzen} T.~J.,
  2017, \mn@doi [\apj] {10.3847/1538-4357/aa88d1}, \href
  {http://adsabs.harvard.edu/abs/2017ApJ...847...64M} {847, 64}

\bibitem[\protect\citeauthoryear{{Moore}, {Aguirre}, {Parsons}, {Jacobs}  \&
  {Pober}}{{Moore} et~al.}{2013}]{Moore2013}
{Moore} D.~F.,  {Aguirre} J.~E.,  {Parsons} A.~R.,  {Jacobs} D.~C.,   {Pober}
  J.~C.,  2013, \mn@doi [\apj] {10.1088/0004-637X/769/2/154}, \href
  {http://adsabs.harvard.edu/abs/2013ApJ...769..154M} {769, 154}

\bibitem[\protect\citeauthoryear{{Moore} et~al.,}{{Moore}
  et~al.}{2017}]{Moore2017}
{Moore} D.~F.,  et~al., 2017, \mn@doi [\apj] {10.3847/1538-4357/836/2/154},
  \href {http://adsabs.harvard.edu/abs/2017ApJ...836..154M} {836, 154}

\bibitem[\protect\citeauthoryear{{Nunhokee} et~al.,}{{Nunhokee}
  et~al.}{2017}]{Nunhokee2017}
{Nunhokee} C.~D.,  et~al., 2017, \mn@doi [\apj] {10.3847/1538-4357/aa8b73},
  \href {http://adsabs.harvard.edu/abs/2017ApJ...848...47N} {848, 47}

\bibitem[\protect\citeauthoryear{Oppermann et~al.,}{Oppermann
  et~al.}{2012}]{Oppermann2012}
Oppermann N.,  et~al., 2012, \mn@doi [\aap] {10.1051/0004-6361/201118526}, 542,
  A93

\bibitem[\protect\citeauthoryear{{Parsons} et~al.,}{{Parsons}
  et~al.}{2014}]{Parsons2014}
{Parsons} A.~R.,  et~al., 2014, \mn@doi [\apj] {10.1088/0004-637X/788/2/106},
  \href {http://adsabs.harvard.edu/abs/2014ApJ...788..106P} {788, 106}

\bibitem[\protect\citeauthoryear{{Patil} et~al.,}{{Patil}
  et~al.}{2017}]{Patil2017}
{Patil} A.~H.,  et~al., 2017, \mn@doi [\apj] {10.3847/1538-4357/aa63e7}, \href
  {http://adsabs.harvard.edu/abs/2017ApJ...838...65P} {838, 65}

\bibitem[\protect\citeauthoryear{{Pen}, {Chang}, {Hirata}, {Peterson}, {Roy},
  {Gupta}, {Odegova}  \& {Sigurdson}}{{Pen} et~al.}{2009}]{Pen2009}
{Pen} U.-L.,  {Chang} T.-C.,  {Hirata} C.~M.,  {Peterson} J.~B.,  {Roy} J.,
  {Gupta} Y.,  {Odegova} J.,   {Sigurdson} K.,  2009, \mn@doi [\mnras]
  {10.1111/j.1365-2966.2009.14980.x}, \href
  {http://adsabs.harvard.edu/abs/2009MNRAS.399..181P} {399, 181}

\bibitem[\protect\citeauthoryear{{Planck Collaboration XIII}}{{Planck
  Collaboration XIII}}{2016}]{PlanckXIII}
{Planck Collaboration XIII} 2016, \aap, 594, A13

\bibitem[\protect\citeauthoryear{{Rybicki} \& {Lightman}}{{Rybicki} \&
  {Lightman}}{1986}]{Rybicki&Lightman1986}
{Rybicki} G.~B.,  {Lightman} A.~P.,  1986, {Radiative Processes in
  Astrophysics}

\bibitem[\protect\citeauthoryear{Santos, Cooray  \& Knox}{Santos
  et~al.}{2005}]{Santos2005}
Santos M.~G.,  Cooray A.,   Knox L.,  2005, \mn@doi [The Astrophysical Journal]
  {10.1086/429857}, 625, 575

\bibitem[\protect\citeauthoryear{{Schnitzeler}, {Katgert}  \& {de
  Bruyn}}{{Schnitzeler} et~al.}{2009}]{Schnitzeler2009}
{Schnitzeler} D.~H.~F.~M.,  {Katgert} P.,   {de Bruyn} A.~G.,  2009, \mn@doi
  [\aap] {10.1051/0004-6361:20078912}, \href
  {http://adsabs.harvard.edu/abs/2009A%26A...494..611S} {494, 611}

\bibitem[\protect\citeauthoryear{{Shaw}, {Sigurdson}, {Sitwell}, {Stebbins}  \&
  {Pen}}{{Shaw} et~al.}{2015}]{Shaw2015}
{Shaw} J.~R.,  {Sigurdson} K.,  {Sitwell} M.,  {Stebbins} A.,   {Pen} U.-L.,
  2015, \mn@doi [\prd] {10.1103/PhysRevD.91.083514}, \href
  {http://adsabs.harvard.edu/abs/2015PhRvD..91h3514S} {91, 083514}

\bibitem[\protect\citeauthoryear{{Singh} et~al.,}{{Singh}
  et~al.}{2017}]{Singh2017}
{Singh} S.,  et~al., 2017, \mn@doi [\apjl] {10.3847/2041-8213/aa831b}, \href
  {http://adsabs.harvard.edu/abs/2017ApJ...845L..12S} {845, L12}

\bibitem[\protect\citeauthoryear{Tegmark}{Tegmark}{1997}]{Tegmark1997}
Tegmark M.,  1997, \mn@doi [\prd] {10.1103/PhysRevD.56.4514}, 56, 4514

\bibitem[\protect\citeauthoryear{{Tingay} et~al.,}{{Tingay}
  et~al.}{2013}]{Tingay2013}
{Tingay} S.~J.,  et~al., 2013, \mn@doi [\pasa] {10.1017/pasa.2012.007}, \href
  {http://adsabs.harvard.edu/abs/2013PASA...30....7T} {30, e007}

\bibitem[\protect\citeauthoryear{{Wang} et~al.,}{{Wang}
  et~al.}{2013}]{Wang2013}
{Wang} J.,  et~al., 2013, \mn@doi [\apj] {10.1088/0004-637X/763/2/90}, \href
  {http://adsabs.harvard.edu/abs/2013ApJ...763...90W} {763, 90}

\bibitem[\protect\citeauthoryear{{Wieringa}, {de Bruyn}, {Jansen}, {Brouw}  \&
  {Katgert}}{{Wieringa} et~al.}{1993}]{Wieringa1993}
{Wieringa} M.~H.,  {de Bruyn} A.~G.,  {Jansen} D.,  {Brouw} W.~N.,   {Katgert}
  P.,  1993, \aap, \href {http://adsabs.harvard.edu/abs/1993A%26A...268..215W}
  {268, 215}

\makeatother
\end{thebibliography}







\bsp	
\label{lastpage}
\end{document}